\documentclass[a4paper,10pt,twoside]{cpc-hepnp}
\usepackage{CJK,upgreek,fancyhdr}
\usepackage{multicol}
\usepackage{graphicx}
\usepackage{booktabs}
\usepackage{amssymb,bm,mathrsfs,bbm,amscd}
\usepackage[tbtags]{amsmath}
\usepackage{lastpage}
\usepackage[justification=centering]{caption}
\usepackage{footnote}

\begin{document}
\begin{CJK*}{GB}{gbsn}

\fancyhead[c]{\small Chinese Physics C~~~Vol. xx, No. x (201x) xxxxxx}
\fancyfoot[C]{\small xxxxxx-\thepage}

\footnotetext[0]{Received xxxx}

\title{Measurement of the $^{232}$Th (n, $\gamma$ )/$^{58}$Ni (n, p) reaction rate ratio in the leakage neutron field of CFBR-II fast burst reactor\thanks{Supported by National Natural Science
Foundation of China (91326104) and Foundation of Key Laboratory of Neutron Physics,China Academy of Engineering Physics(2015AA01,2014AA01) }}

\author{%
 %     Wang Qiang(王强)$^{1)}$ \email{johnywon@qq.com}%
 Qiang Wang(王强)$^{1)}$ \email{johnywon@qq.com}%
\quad Li-Na Zeng (曾丽娜)
\quad Zi-Hui Ai (艾自辉)
\quad Chun Zheng （郑春）
\quad Jian Gong （龚建）
%}lab1
}
\maketitle

\address{%
Institute of Nuclear Physics and Chemistry, China Academy of Engineering Physics, 621900, Miyang, China\\

}

\begin{abstract}
A ThO$_{2}$ sample and a nickel activation foil were irradiated in the leakage neutron field of CFBR-II reactor. The activities of the activation products were measured after irradiation to obtain the reaction rates. The normalized reaction rates were also calculated based on the ENDF/B-VII.1, CENDL-3.1, JENDL-4.0, BROND-2.2 databases. The experimental reaction rate ratio is 4.37 with an uncertainty of 3.9\% which is coincident with each of the ratios calculated based on the ENDFB-VII. 1, JENDL-4.0, BROND-2.2 databases, but is 11.2\% larger than that based on CENDL-3.1 database.
\end{abstract}



\footnotetext[0]{\hspace*{-3mm}\raisebox{0.3ex}{$\scriptstyle\copyright$}2013
Chinese Physical Society and the Institute of High Energy Physics
of the Chinese Academy of Sciences and the Institute
of Modern Physics of the Chinese Academy of Sciences and IOP Publishing Ltd}%

\begin{multicols}{2}

\section{Introduction}

Thorium/Uranium fuel cycle is an important research content of the fourth generation of reactors \citep{lab1}.  $^{232}$Thorium’s abundance of Thorium isotopes in nature is almost 100 \% . When neutrons are absorbent by $^{232}$Th nuclides, $^{232}$Th (n,$\gamma$) reactions will take place. After two cascade decays，the capture reaction product $^{233}$Th will transform into fissionable $^{233}$U nuclides entirely. This so called Th/U fuel cycle is not a new concept but a new trend to researchers in recent years.

Given that there are some non-negligible differences between the evaluation data and lack of macroscopic neutronics experimental data about $^{232}$Th relatively \citep{lab2}, it is necessary to carry out experiments to measure and to validate the nuclear data in varies neutron field. Recently, to reevaluate the accuracy of reactor physics parameters, Masao Yamanaka, Cheol Ho Pyeon, Takahiro Yagi et al. have irradiated Thorium foils in the Thorium-Loaded Accelerator-Driven System Experiments at Kyoto University Critical Assembly, and then the 311.9keV characteristic $\gamma$ rays emitted from the foils were counted to obtain the $^{232}$Th (n,$\gamma$) reaction rate, the statistical errors are 3\% for the rates \citep{lab3}.In the 90s of last century, Institute of Applied Physics of China Academy of Science has established a Thorium-Uranium zero power facility ,and carried out many critical experiments furthermore \citep{lab1,lab4}. In 2013, Yang Yiwei, Liu Rong and Yan Xiaosong measured the reaction rate of $^{232}$Th(n,$\gamma$) which the ThO$_2$ samples were irradiated by 14MeV neutrons at the different locations in a polyethylene ball and the uncertainty is about 6\% typically \citep{lab5}. In 2014 our group has also measured the capture reaction rate of $^{232}$Th in the leakage radiation field of a columnar Uranium fast sub-critical assembly, and we found that the capture reaction rate calculated based CENDL-3.1 is 18.5\% smaller than that of the experiment result \citep{lab6}.

Generally, activation method is adopted in reaction rates measurement. The detection efficiency of the $\gamma$ rays has the largest uncertainty for the reaction rate measurement which is about 3\% typically \citep{lab7}. It is commonly to use nickel foil to measure the neutron flux for fast neutrons. In the standard ASTM E264---02 , it can be seen that general practice indicates that the reaction rate can be determined with a bias of 3\%\citep{lab8}.To obtain the information of the capture reaction rate and the neutron flux in a fast neutron field,a ThO$_2$ sample and a nickel foil were irradiated by the leakage neutrons from CFBR-II(Chinese Fast Burst Reactor-II) reactor and the $\gamma$ rays from the samples were counted after irradiation to obtain the $^{232}$Th (n,$\gamma$) / $^{58}$Ni (n, p) reaction rate ratio (hereinafter referred to as the reaction rate ratio) in this study. The experimental reaction rate ratio is 4.37 which is coincident with each of the ratios calculated based on ENDF/B-VII.1, JENDL-4.0 and BROND-2.2 databases, but is 11.2\% larger than that based on CENDL-3.1 database.

With an alternative method to calibrate the HPGe $\gamma$ spectrometer, the uncertainty of the experimental reaction rates ratio of $^{232}$Th (n,$\gamma$) and  $^{58}$Ni (n, p) is estimated to be 3.9\%.It is believed that if the uncertainties of the standard sources employed for $\gamma$ spectrometer's calibration are both reduced to about 1\% respectively, the reaction ratio could reach a lower level around 2\%.

\section{Principles and methods}
\subsection{Basic principles}
After the irradiation of the ThO$_{2}$ sample and the nickel foil by the leakage neutrons, the activities of the activation products can be obtained based on the $\gamma$ rays’ counts, then the reaction rate ratio can be obtained based on these data.

\begin{eqnarray}
\label{eq1}
^{232}\textrm{Th}+n\xrightarrow[]{}^{233}\textrm{Th}\xrightarrow[100\%]{\beta^-}^{233}\textrm{Pa}\xrightarrow[100\%]{\beta^-}^{233}\textrm{U}.
\end{eqnarray}

As can be seen from (\ref{eq1}), $^{233}$U nuclides are created after two beta decays from the $^{233}$Th nuclides produced by $^{232}$Th (n,$\gamma$) reaction.The number of the $^{233}$Th nuclides at the end of the irradiation
\begin{eqnarray}
\label{eq2}
N_{1}=\frac{N_0 \bar\sigma \phi}{\lambda_1-\bar\sigma\phi}\left ( e^{-\bar\sigma\phi t_0}-e^{-\lambda_1t_0} \right ).
\end{eqnarray}

\noindent
where $N_{0}$ is the number of $^{232}$Th nuclides in the ThO$_{2}$ sample, ${\bar\sigma}$ is the averaged capture cross section of $^{232}$Th ,${\phi} $ is the neutron flux,$t_{0}$ is the time interval of the irradiation,and $\lambda_{1}$  is the decay constant of $^{233}$Th.

 The half life is 22.3min for $^{233}$Th and 26.975d for $^{233}$Pa.As it is difficult to measure the activity of $^{233}$Th, we introduce an alternative method which is to measure the activity of $^{233}$Pa to obtain the capture reaction rate of $^{232}$Th.

The number of $^{233}$Pa nuclides at the beginning of the activity measurement for $^{233}$Pa

\begin{eqnarray}
\label{eq3}
N_2 &=&\frac{N_0 \bar\sigma \phi \lambda_1\left ( e^{-\bar\sigma\phi t_0}-e^{-\lambda_1t_0} \right )\left ( e^{-\lambda_1t_1}-e^{-\lambda_2t_1} \right )}{\left(\lambda_1-\bar\sigma\phi\right)\left(\lambda_2-\lambda_1\right)}
\nonumber\\
&&+N_0 \bar\sigma \phi\left ( h_1e^{-\bar\sigma\phi t_0}+h_2e^{-\lambda_1t_0}+h_3e^{-\lambda_2t_0} \right )e^{-\lambda_2t_1}
\nonumber\\
%&&
%\nonumber\\
&=&N_0\bar\sigma\phi R\left (t \right ).
\end{eqnarray}

\noindent
where $\lambda_2$ is the decay constant of $^{233}$Pa , $t_{1}$ is the waiting time, $R\left (t \right )$ is the time factor, and

\begin{displaymath}
h_1=\frac{\lambda_1}{\left (\lambda_1-\bar\sigma\phi  \right )\left (\lambda_2-\bar\sigma\phi   \right )}.
\end{displaymath}

\begin{displaymath}
h_2=\frac{\lambda_1}{\left (\lambda_1-\bar\sigma\phi  \right )\left (\lambda_1-\lambda_2   \right )}.
\end{displaymath}

\begin{displaymath}
h_3=\frac{\lambda_1}{\left (\lambda_2-\bar\sigma\phi  \right )\left (\lambda_2-\lambda_1  \right )}.
\end{displaymath}

Then the capture reaction rate

\begin{eqnarray}
\label{eq4}
R_{\rm{Th}}&=&\bar\sigma\phi=\frac{N_2}{N_0 R\left (t \right )}.
\end{eqnarray}

The $^{58}$Ni (n,p) $^{58}$Co reaction rate could be obtained by measuring the activity of $^{58}$Co similarly.

\subsection{Methods for $\gamma$ spectroscopy}

The less disturbed 311.9keV $\gamma$ rays of  $^{233}$Pa and 810.8keV $\gamma$ rays of $^{58}$Co were measured to obtain the reaction rate ratio.To reduce the uncertainty of the measurement, a $^{237}$Np and a $^{152}$Eu source were employed to calibrate the $\gamma$ spectrometer to acquire the detection efficiency of 311.9keV and 810.8keV $\gamma$ rays separately.

\begin{eqnarray}
\label{eq5}
^{237}\textrm{Np}\xrightarrow[100\%]{\alpha}^{233}\textrm{Pa}\xrightarrow[100\%]{\beta^-}^{233}\textrm{U}.
\end{eqnarray}

As illustrated in (~\ref{eq5}), the only daughter of $^{237}$Np is $^{233}$Pa.For that the half life of $^{233}$Pa is far shorter than that of $^{237}$Np, after a period by the $^{237}$Np source was prepared, the activity of $^{233}$Pa in the $^{237}$Np source will actually equal to the activity of $^{237}$Np.That is, a $^{237}$Np source could be use as a $^{233}$Pa $\gamma$ source. Generally, $^{237}$Np source is also a strong ${\alpha}$ emitter, and the ${\alpha}$ rays can be accurately counted to obtain the activity with an even smaller uncertainty. So, a $^{237}$Np source, actually a $^{233}$Pa source was employed to calibrate the HPGe spectrometer to obtain the accurate detection efficiency of $\gamma$ rays of 311.9keV.

Thus the capture reaction rate of $^{232}$Th can be acquired experimentally:

\begin{eqnarray}
\label{eq6}
R_{\rm{Th}}&=&\bar\sigma\phi=\frac{N_2}{N_0 R\left (t \right )}=\frac{A_{\rm{Pa}}}{\lambda_2 N_0 R\left (t \right )}
\nonumber\\
&=&\frac{C}{b \epsilon }\frac{\lambda_2}{f \left (1-e^{-\lambda_2 t_2}\right ) }\delta \times \frac{1}{\lambda_2 N_0 R\left (t \right )}
\nonumber\\
&=& \frac{C}{b }\frac{\lambda_2}{f \left (1-e^{-\lambda_2 t_2}\right ) }\delta \times
\nonumber\\
&&\frac{1}{\frac{C_{\rm{Np}}}{A_{\rm{Np}} b \delta\rm{_{Np}}F_{\rm{Np}}\left ( t \right )}}\times \frac{1}{\lambda_2 N_0 R\left (t \right )}
\nonumber\\
&=&\frac{ C \lambda_{\rm{Pa}} F \left ( t \right ) \delta } { N_{\rm{A}} \frac{m_{i}}{M_{\rm{ThO_2}}}} \times \frac{A_{\rm{Np}}F_{\rm{Np}}\left ( t \right )\delta_{\rm{Np}}}{C_{\rm{Np}}}.
\end{eqnarray}

\noindent
where \emph{A$_{\rm{Pa}}$} is the activity of $^{233}$Pa ,$C$ is the net counts of the photoelectric peak of 311.9keV $\gamma$ rays, $b$ is the branching ratio, $\epsilon$ is the detection efficiency , $f$ is the ratio of live time and real time, $t_{2}$ is the live time, ${\delta}$ is the correction factor and $F\left (t \right )$ is the experimental time factor,$m_{i}$ is the mass of the sample, $M_{\rm{ThO_2}}$ is the mass of 1 mol ThO$_{2}$, $N_{\rm{A}}$ is Avogadro constant. The symbols each with a subscript of 'Np' mean that they were acquired in the 311.9keV $\gamma$ rays' detection efficiency calibration,$\delta_{\rm{Np}}$ is the the self-absorption correction factor for $^{237}$Np source.And

\begin{displaymath}
\delta = \delta_1 \times \delta_2.
\end{displaymath}

\noindent
where $\delta_1$ is the source area correction factor,$\delta_2$ is the self-absorption correction factor.

Furthermore, the energy of the 810.8keV $\gamma$ rays emitted from a $^{58}$Co source is almost the same as that of the 810.5keV $\gamma$ rays emitted from a $^{152}$Eu $\gamma$ source. Since we do not have a proper $^{58}$Co source to calibrate the HPGe spectrometer, the detection efficiency of 810.5keV $\gamma$ rays which was calibrated by a standard $^{152}$Eu source is used as the efficiency of 810.8keV $\gamma$ rays instead ,and it is believed the difference of the efficiencies between 810.8keV and 810.5keV could be ignored. Thus, $^{58}$Ni (n, p) reaction rate

\begin{eqnarray}
\label{eq7}
R_{\rm{Ni}}&=& \frac{C_{\rm{Co}}}{b_{\rm{Co}}}\frac{\lambda_{\rm{Co}} e^{\lambda_{\rm{Co}} t_{0}}}{f_{\rm{Co}}\left (1-e^{-\lambda_{\rm{Co}} t_{1,\rm{Co}}}  \right )\left (1-e^{-\lambda_{\rm{Co}} t_{1,\rm{Co}}}  \right )}
\nonumber\\
&&\frac{1}{N_{0,\rm{Ni}}}\delta_{\rm{Co}}\frac{1}{\frac{C_{\rm{Eu}}}{A_{\rm{Eu}} b_{\rm{Eu}} \delta_{\rm{Eu}}F_{\rm{Eu}}\left ( t \right )}}
\nonumber\\
&=&\frac{ C_{\rm{Co}} \lambda_{\rm{Co}} F_{\rm{Co}} \left ( t \right ) \delta_{\rm{Co}} } {b_{\rm{Co}} p N_{\rm{A}} \frac{m_{\rm{Ni}}}{M_{\rm{Ni}}}} \times \frac{A_{\rm{Eu}}F_{\rm{Eu}}\left ( t \right )\delta_{\rm{Eu}}b_{\rm{Eu}}}{C_{\rm{Eu}}}.
\end{eqnarray}

\noindent
where, $p$ is the concentration of $^{58}$Ni in natural nickel,$m_{\rm{Ni}}$ is the mass of the nickel foil, $M_{\rm{Ni}}$ is the mass of 1 mol nickel.The symbols with the subscript of 'Co' mean that they were acquired in the measurement for the activation foil ,and subscript of 'Eu' mean that they were acquired in the 810.5keV $\gamma$ rays' detection efficiency calibration.

%%20161101

\section{Experiments and results}
\subsection{CFBR-II reactor and the samples}

CFBR-II reactor was used as the neutron source for irradiation. The reactor can be operated in a steady power state, and the neutrons have a near fission energy spectrum. The average energy of the neutrons out of the core is 1.23MeV \citep{lab9}.

The powder has a content of 99.9 \% ThO$_{2}$ was used to prepare the irradiated sample. 0.09899 gram powder which was weighted by a precision balance was compacted evenly into a D15mm disk and then placed in an aluminum case. A D10mm, 0.13gram high purity nickel foil was fixed onto the surface of the aluminum case next to the ThO$_{2}$ sample for irradiation.

\subsection{Irradiation and radioactivity measurements}

The ThO$_{2}$ sample and the nickel foil were fixed on the surface of CFBR-II reactor for irradiation. CFBR-II reactor was operated at the power level of 300W to irradiate the samples for 3530s.

After irradiation, the $\gamma$ rays from the ThO$_{2}$ sample were counted by a HPGe spectrometer from PGT(Princeton Gamma-Tech, Inc.) at the position of 6.5cm from the detector's upper surface , however, the nickel foil was placed on the upper surface of the detector for measuring. Then the activities of $^{233}$Pa and $^{58}$Co were acquired based on the counts of the $\gamma$ rays of 311.9keV and 810.8keV, therefore, the reaction rate ratio was achieved.

\begin{center}
\includegraphics[width=8cm]{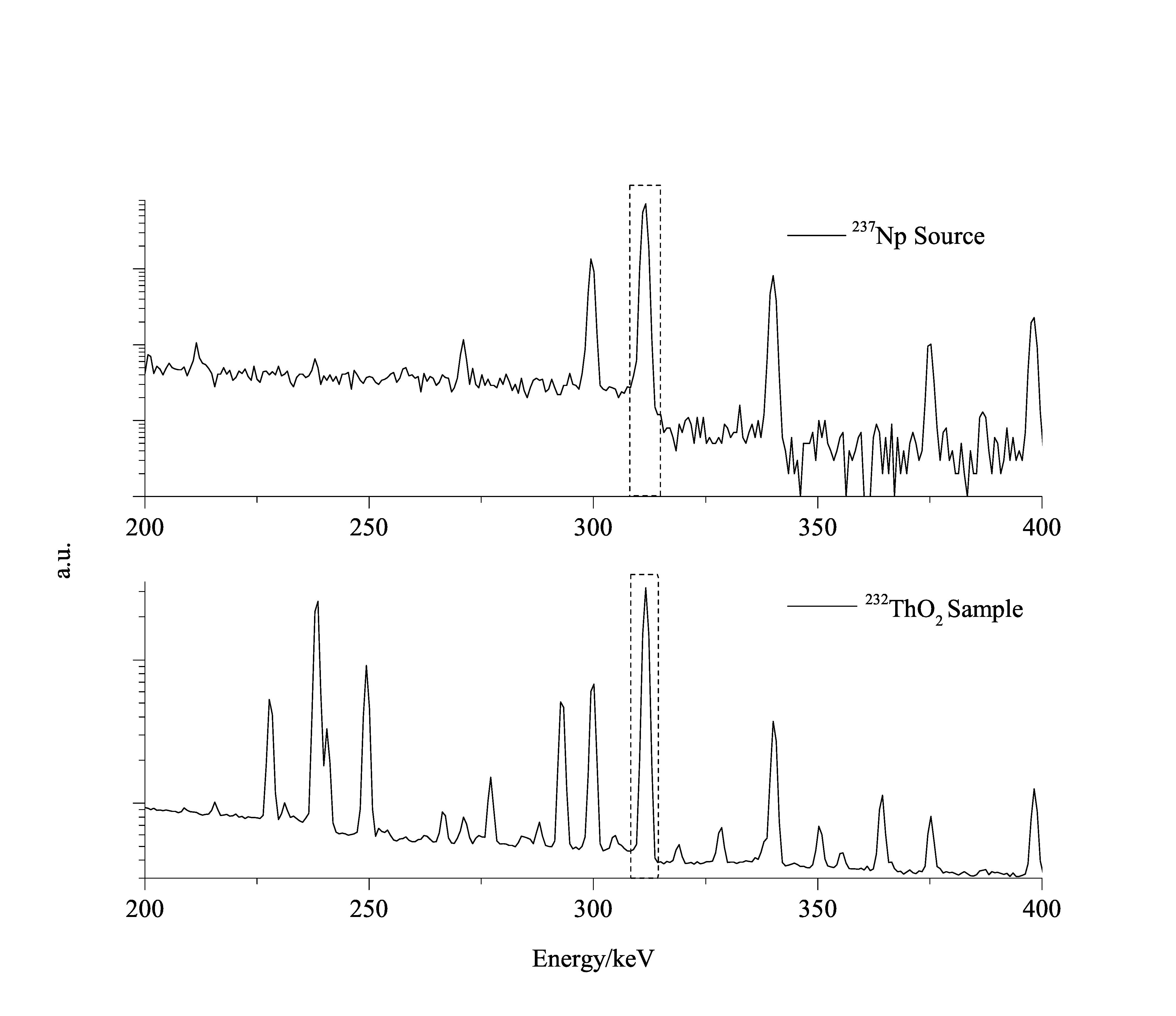}
%cpcf2.eps,
\figcaption{\label{fig1} The $\gamma$ spectra of $^{237}$Np source and the ThO$_{2}$ sample after irradaition }
\end{center}

\begin{center}
\includegraphics[width=8cm]{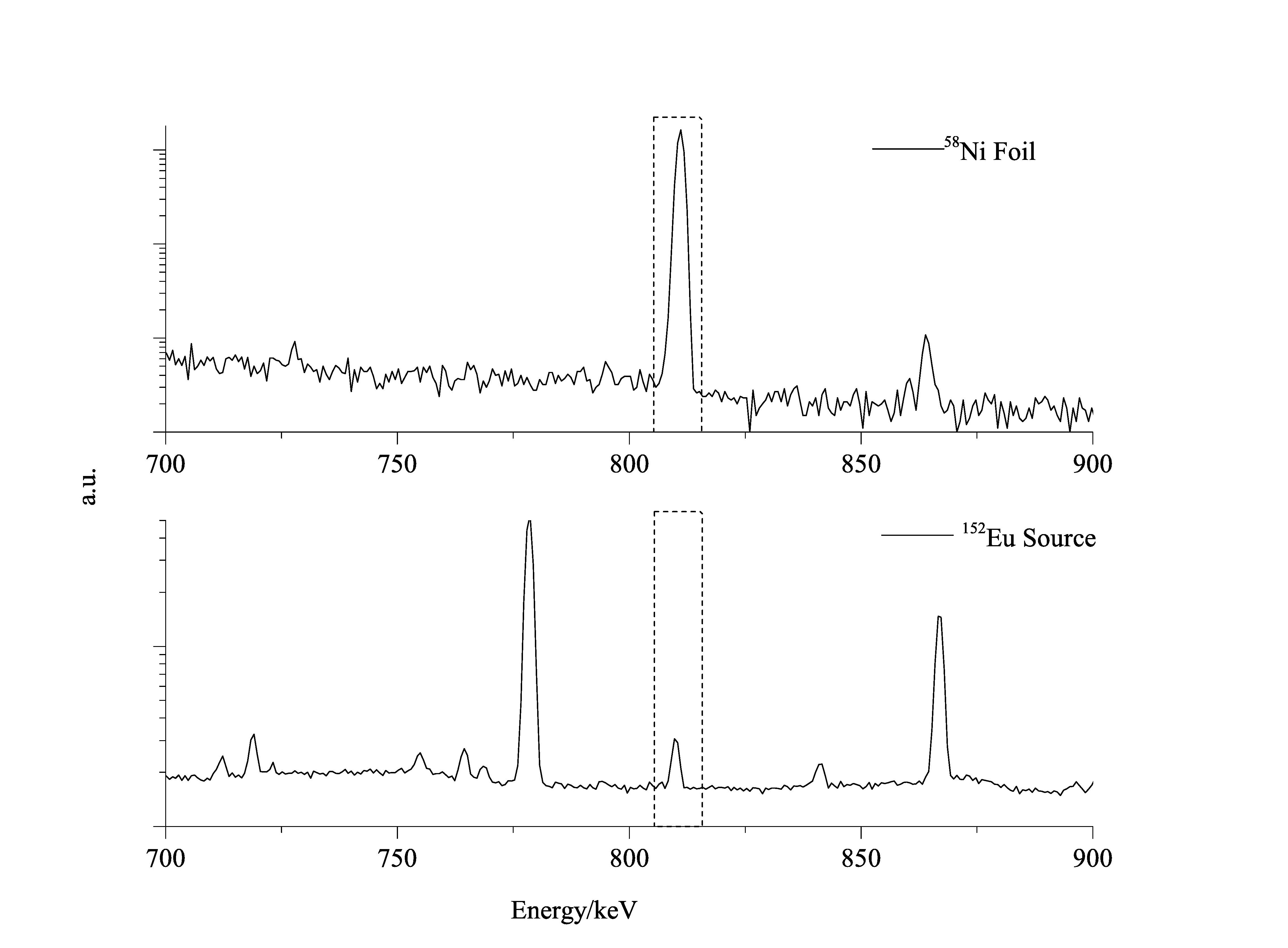}
%cpcf2.eps,
\figcaption{\label{fig2} The $\gamma$ spectra of $^{237}$Eu source and the nickel foil after irradaition }
\end{center}

Fig.~(\ref{fig1}) and Fig.~(\ref{fig2}) are the measured $\gamma$ spectra.In Fig.~(\ref{fig1}),the corresponding energies of peaks in the dash boxes are both 311.9keV. In Fig.~(\ref{fig2}),the corresponding energy of peak in the upper dash box is 810.8keV and in the lower dash box is 810.5keV.

\subsection{Data processing and experiment result}

Nuclear data used for the data processing are exported from ENDF/B-VII.1 database. The half life of $^{233}$Pa is 26.975d, branching ratios of 311.9keV  $\gamma$ ray is 0.385, half life of $^{58}$Co is 70.86d, branching ratio of 810.8keV $\gamma$ ray is 0.9945.Ratios of the live time and real time were exported directly from the record of the spectrometer with a typical value of 0.973. The reaction rate is very small, so we consider $e^{-\bar\sigma\phi t_0}$  in the time factor as 1.

Monte Carlo method is employed to calculate the self absorption correction factors and the source-area correction factors like that in Ref.~\citep{lab6}. For ThO$_{2}$ sample they are 0.989 and 1.005.For nickel foil they are 0.987 and 1.01.Thus the ratio of $^{232}$Th (n,$\gamma$) and $^{58}$Ni (n, p) reaction rate can be obtained and is listed in Table~\ref{tab1}.

\begin{center}
\tabcaption{ \label{tab1}  Ratio of the reaction rates}
\footnotesize
\begin{tabular*}{80mm}{c@{\extracolsep{\fill}}ccc}
\toprule data source & reaction rate ratio \\
\hline
Experimental & 4.37 \\
Calculated(ENDF/B-VII.1) & 4.34\\
Calculated(CENDL-3.1) & 3.88\\
Calculated(JENDL-4.0) & 4.38\\
Calculated(BROND2.2) & 4.26\\
\bottomrule
\end{tabular*}
\vspace{0mm}
\end{center}
\vspace{0mm}
%%\end{multicols}
%%\begin{multicols}{2}

\section{Analysis and discussion}
\subsection{Uncertainty analysis}

\subsubsection{Uncertainty analysis for the $^{232}$Th (n,$\gamma$) reaction rate }

 The uncertainty of the $^{232}$Th (n,$\gamma$) reaction rate can be analysed based on Eq. \ref{eq6}. Main components to the uncertainty of capture reaction rate of $^{232}$Th (n,$\gamma$) are listed in Table 2. The activity of the standard $^{237}$Np source was previously measured by an $\alpha$ spectrometer and the uncertainty was estimated to be 0.025.The uncertainty of the correction factor  is assumed to be itself which is 0.006.The smaller uncertainties are all ignored.

\begin{center}
\tabcaption{ \label{tab2}Uncertainty components of the $^{232}$Th (n,$\gamma$) reaction rate}
\footnotesize
\begin{tabular*}{80mm}{c@{\extracolsep{\fill}}ccc}
\toprule Component & Evaluation method & Uncertainty  \\
\hline
$u\left(C \right)$ & A & 0.01 \\
$u\left(C_{\rm{Np}} \right)$ & A & 0.01 \\
$u\left(\delta \right)$ & B & 0.006 \\
$u\left(A_{\rm{Np}} \right)$ & B & 0.025 \\
\bottomrule
\end{tabular*}
\vspace{0mm}
\end{center}
\vspace{0mm}
%%\end{multicols}
%%\begin{multicols}{2}

The items in table~\ref{tab2} are not relevant, whereby the relative standard uncertainty
\begin{eqnarray}
\label{eq8}
u_c = \sqrt{\sum_{i}^{}\left ( \frac{\partial {R_{\rm{Th}}}}{\partial x_i} \right )^2u^2\left ( x_i \right )}=0.029.
\end{eqnarray}

\subsubsection{Uncertainty analysis for the $^{58}$Ni(n,p) reaction rate}

The uncertainty of the $^{58}$Ni (n,p) reaction rate can be analysed based on Eq.~\ref{eq7}. The activity's uncertainty of the standard $^{152}$Eu source is exported from its manual as 0.02. The uncertainty of the correction factor  is assumed to be itself which is 0.003. The uncertainty of the branching ratio of 810.5keV $\gamma$ ray is exported from ENDF / B-VII.1 and it is 0.0096.The smaller uncertainties are all ignored.

\begin{center}
\tabcaption{ \label{tab3}  Uncertainty components of the $^{58}$Ni (n,p) reaction rate}
\footnotesize
\begin{tabular*}{80mm}{c@{\extracolsep{\fill}}ccc}
\toprule Component	 & Evaluation method & Uncertainty  \\
\hline
$u\left(C_{\rm{Ni}} \right)$ & A & 0.01 \\
$u\left(C_{\rm{Eu}} \right)$ & A & 0.01 \\
$u\left(\delta_{\rm{Co}} \right)$ & B & 0.003 \\
$u\left(A_{\rm{Eu}} \right)$ & B & 0.02 \\
$u\left(b_{\rm{Eu}} \right)$ & A & 0.0096 \\
\bottomrule
\end{tabular*}
\vspace{0mm}
\end{center}
\vspace{0mm}
%%\end{multicols}
%%\begin{multicols}{2}
The items in table~\ref{tab3} are not relevant, whereby the relative standard uncertainty can be calculated as 2.6\% like Eq.~\ref{eq8}.

The two reaction rates are not related, thus the synthesis relative uncertainty of the reaction rate ratio can be estimated as 3.9\% ,and the uncertainties of the standard sources' activities are still the main sources for the total uncertainty.
\subsection{Comparison of the results from calculation and measurement}

The group cross sections were prepared based on ENDF/B-VII.1, CENDL-3.1, JENDL-4.0 and BROND-2.2 databases by the weighting function of the free software package JANIS4.0 \citep{lab10}. To achieve the calculated reaction rates, the convolution of the neutron spectrum and the group cross sections was computed. The neutron spectrum for calculation was measured by a $^6$Li sandwiched spectrometer combined with radioactive foils before. Therefore, as listed in Table~\ref{tab1}, the calculated ratio of the $^{232}$Th (n,$\gamma$) and $^{58}$Ni(n,p) reaction rates was obtained .

There is a small discrepancy between the experiment result and each of the reaction rate ratios calculated based on ENDF/B-VII.1, JENDL-4.0 and BROND-2.2 databases. But the maximum difference occurs when the CENDL-3.1 database is used for the calculated reaction ration's compution, and the relative deviation is 11.2\%.

We have carried out a similar experiment in the leakage neutron field of a columnar critical assembly previously and found that the average cross section of $^{232}$Th (n,$\gamma$) calculated based on the CENDL-3.1 database is 18.5\% smaller than that by measurement. The result of the comparison is similar to this experiment. Taking into account that the reaction rate ratio reflects the ratio of average cross-sections, we believe that there are non-negligible differences between the databases in fast neutron energy region, and especially more work should be carried out on the data of $^{232}$Th (n,$\gamma$) reaction in CENDL-3.1 database.
\section{Conclusions}

A ThO$_2$ sample and a nickel foil were irradiated in the leakage neutron field of CFBR-II reactor, then the 311.9keV and the 810.8keV $\gamma$ rays emitted from the sample and the foil were counted respectively to obtain the ratio of $^{232}$Th (n,$\gamma$) and $^{58}$Ni(n,p) reaction rates. The ratio measured is 4.37 with an uncertainty of 3.9\%.There is a good coincidence between the experiment result and each of the calculated reaction rate ratios based on some evaluated databases such as JENDL-4.0,BROND-2.2 and ENDF/B-VII.1,but a largest deviation occurs as 11.2\% when the cross sections from CENDL-3.1 are used for calculation.

In order to obtain a less uncertainty for the reaction rate ratio, it is helpful to utilize standard sources with low uncertainties to calibrate the HPGe spectrometer. Providing that the uncertainties are 1\% for both of the $^{152}$Eu and the $^{237}$Np sources, generally not a great problem, the uncertainty of the reaction rate ratio would reach a lower level which is about 2\%.
\\

\acknowledgments{The authors would like to thank the CFBR-II crew for the irradiation and thank Professor Jian-Sheng Li for his enthusiastic and helpful discussions.
}

\end{multicols}

\vspace{-1mm}
\centerline{\rule{80mm}{0.1pt}}
\vspace{2mm}

\begin{multicols}{2}

\end{multicols}

\newpage

\begin{multicols}{2}
\end{multicols}

\clearpage
\end{CJK*}
\end{document}